\documentclass[aps,prl,twocolumn,groupedaddress]{revtex4}

\usepackage{graphicx}

\begin{document}

\title{Superfluid response in electron-doped cuprate superconductors}
\author{H. G. Luo}
\author{T. Xiang}
\affiliation{Institute of Theoretical Physics and
Interdisciplinary Center of Theoretical Studies, Chinese Academy
of Sciences, P. O. Box 2735, Beijing 100080, China}

\begin{abstract}

We propose a weakly coupled two-band model with $d_{x^2-y^2}$
pairing symmetry to account for the anomalous temperature dependence
of superfluid density $\rho_s$ in electron-doped cuprate
superconductors. This model gives a unified explanation to the
presence of an upward curvature in $\rho_s$ near $T_c$ and a weak
temperature dependence of $\rho_s$ in low temperatures. Our work
resolves a discrepancy in the interpretation of different
experimental measurements and suggests that the pairing in
electron-doped cuprates has predominately $d_{x^2-y^2}$ symmetry in
the whole doping range.

\end{abstract}

\maketitle


Identification of pairing symmetry has been an important issue in
the investigation of high-Tc superconductivity. For hole-doped
cuprate superconductors, it is commonly accepted that the pairing
order parameter has d$_{x^{2}-y^{2}}$-wave symmetry
\cite{tsuei00a}. However, for electron-doped cuprate
superconductors, no consensus has been reached on the pairing
symmetry. A number of experiments, including the angle resolved
photoemission (ARPES) \cite{sato01,armitage01}, the Raman
spectroscopy \cite {blumberg02}, and the phase-sensitive
measurements \cite{tsuei00b,chesca03}, suggested that the
electron-doped superconductors also have d$_{x^{2}-y^{2}}$ -wave
symmetry. However, the results revealed by other experiments are
controversial \cite{alff99,kokales00,prozorov00,prozorov04,kim03}.
In particular, the magnetic penetration depth data measured by
Kokales {\it et al.} \cite{kokales00} and by Prozorov {\it et al.}
\cite {prozorov00,prozorov04} showed that the low temperature
superfluid density of electron-doped superconductors varies
quadratically with temperature in the whole range of doping, in
agreement with the theoretical prediction for a d-wave
superconductor with impurity scattering. However, the experimental
data published by Kim {\it et al.} \cite{kim03} suggested that
there is a d- to anisotropic s-wave transition across the optimal
doping. For optimal and overdoped samples, they found that the low
temperature superfluid density exhibits an exponential temperature
dependence, in favor of an anisotropic s-wave pairing state.

The above discrepancy indicates that low lying quasiparticle
excitations in electron-doped cuprates behave quite differently than
in hole doped ones. To resolve the discrepancy, a thorough
understanding of the electronic structure of electron-doped
materials is desired. In this regard, the doping evolution of the
Fermi surface (FS) revealed by the ARPES of Nd$_{2-x}$Ce$_x$CuO$_4$
(NCCO) \cite{armitage02} is of great interest. At low doping, a
small FS pocket first appears around $(\pi ,0)$, in contrast to the
hole doping case where the low-lying states are centered around
$(\pi /2,\pi /2)$\cite{yoshida03}. By further doping, another pocket
begins to form around $(\pi /2,\pi /2)$. The presence of the two
separate FS pockets may result from the band folding effect induced
by the antiferromagnetic correlations\cite{Matsui04,yuan04}. It may
also be a manifestation of the lower and upper Hubbard
bands\cite{kusko02}. At the mean field level, the theoretical
calculations indicate that these two FS pockets can be {\it
effectively} described as a two-band system \cite{yuan04,kusko02}.
This two-band scenario is consistent with the conjecture made by a
number of groups \cite{wang91,jiang94,fournier97} on the existence
of two kinds of charge carriers in electron-doped materials.

The interplay between the above mentioned two bands can affect
significantly the behavior of superconducting quasiparticles. A
generic feature of a weakly coupled two-bands system, as first
pointed out by Xiang and Wheatley \cite{xiang96}, is the presence of
an upward curvature in the temperature dependence of superfluid
density $\rho_s$ near $T_c$. This intrinsic upward curvature in the
superfluid density has indeed been observed in electron-doped
materials by a number of experimental groups
\cite{kim03,skinta02a,skinta02b,pronin03}. Not only does it lend
further support to the two-band picture, but also sheds light on the
understanding of various controversial experimental observations.

In this paper, we propose to use a two-band BCS-like model with
d$_{x^2-y^2}$-wave pairing symmetry to account for the low energy
electromagnetic response of superconducting quasiparticles in
electron-doped materials. This model, as will be shown later,
captures the main features of quasiparticle excitations in the
superconducting state and gives a unified account for the
experimental data. Our result suggests that the superconducting
pairing in electron-doped materials is governed by the same
mechanism as in hole-doped ones, although their phase diagrams
look asymmetric.

The two-band model we study is defined by
\begin{eqnarray}
H&=&\sum_{ik\sigma }\xi _{ik}c_{ik\sigma }^{\dagger }c_{ik\sigma
}+\sum_{ikk^{\prime }}V_{ikk^{\prime }}c_{ik^{\prime }\uparrow
}^{\dagger }c_{i-k^{\prime }\downarrow }^{\dagger }c_{i-k\uparrow
}c_{ik\downarrow }
\nonumber \\
&&+\sum_{kk^{\prime }}\left( V_{3kk^{\prime }}c_{1k^{\prime
}\uparrow }^{\dagger }c_{1-k^{\prime }\downarrow }^{\dagger
}c_{2-k\uparrow }c_{2k\downarrow } + h.c.\right),  \label{ham1}
\end{eqnarray}
where $i=1,2$ represents the band around $\left( \pi ,0\right) $
and that around $\left( \pi /2,\pi /2\right) $, respectively.
$c_{1k\sigma }$ and $ c_{2k\sigma }$ are the corresponding
electron operators. $V_{1kk^{\prime }}$ and $V_{2kk^{\prime }}$
are the reduced pairing potentials for the two bands. $
V_{3kk^{\prime }}$ is the interband pair interaction. This model
has also been used to describe superconducting properties of the
two-band superconductor MgB$_2$\cite{nakai02}. In MgB$_2$, the
interband coupling is weak since the two relevant bands have
different parity symmetry \cite{mazin02}. In the present case, the
interband coupling is also weak since the strong antiferromagnetic
fluctuations do not couple the first band with the second one in
electron-doped cuprates.

In electron-doped materials, the superconductivity occurs at much
higher doping than in hole doped ones. However, as shown by the
ARPES experiments, the appearance of the superconducting phase
coincides with the appearance of the second band at the Fermi
level. This reveals a close resemblance between electron- and
hole- doped materials. It suggests that it is the interaction
driving the second band to superconduct that leads the whole
system to superconduct in electron-doped materials, and that the
pairing potential $V_{2kk^{\prime }}$ has predominantly
d$_{x^2-y^2}$ symmetry, resembling the hole doped case.
$V_{1kk^{\prime }}$ can in principle be different to
$V_{2kk^{\prime }}$. However, if pairing in the first band is
originated from the same mechanism as the second band or induced
by the second band by the proximity effect, $V_{1kk^{\prime }}$
should most probably have d$_{x^2-y^2}$ symmetry.

In the calculations below, we assume that $V_{ikk^{\prime }}$
($i=1,2,3$) can all be factorized: $V_{1kk^{\prime }}=g_{1}\gamma
_{k}\gamma _{k^{\prime }}$ , $V_{2kk^{\prime }}=g_{2}\gamma
_{k}\gamma _{k^{\prime }}$ and $ V_{3kk^{\prime }}=g_{3}\gamma
_{k}\gamma _{k^{\prime }}$, where $g_{1}$, $ g_{2}$, and $g_{3}$ are
the corresponding coupling constants, $\gamma_{1k} = \gamma _{2k} =
\gamma _{k} = \cos k_{x}-\cos k_{y}$ is the d$_{x^2-y^2}$-wave
pairing function. Here we have implicitly assumed that the first
band has the same pairing symmetry as the second one. This
assumption can in fact be relaxed. The qualitative conclusion draw
below does not depend much on the detailed form of the pairing
function for the first band near $(\pi,0)$, provided there are no
gap nodes on the FS of this band.

Taking the BCS mean field approximation, the interaction between the
two bands is decoupled. It is straightforward to show that the
quasiparticle eigenspectrum of the \textit{i}'th band is given by
the following expression $E_{ik}=\sqrt{\xi _{ik}^{2}+\Delta
_{i}^{2}\gamma _{k}^{2}}$, where $\Delta _{i}$ is the gap amplitude
of the {\it i}'th band. They are determined by the following coupled
gap equations $ \Delta _{1} = \sum_{k}\gamma _{k}( g_{1}\langle
c_{1-k\downarrow }c_{1k\uparrow }\rangle +g_{3}\langle
c_{2-k\downarrow }c_{2k\uparrow }\rangle)$ and $\Delta _{2} =
\sum_{k}\gamma _{k}( g_{2}\langle c_{2-k\downarrow }c_{2k\uparrow
}\rangle +g_{3}\langle c_{1-k\downarrow }c_{1k\uparrow }\rangle )$,
where $\langle \cdots \rangle$ denotes thermal average.

The above expression of $E_{ik}$ indicates that there are gap nodes
in the quasiparticle excitations of the second band. However, there
is a finite excitation gap in the first band since the nodal lines
of $\gamma_k$ do not intersect with the FS of that band if the
system is not heavily overdoped. Therefore, as far as thermal
excitations are concerned, the first band behaves as in a s-wave
superconductor, although the pairing is of $d_{x^2-y^2}$ symmetry.
This indicates that the superconducting state of electron-doped
cuprates is actually a mixture of d-wave and s-wave-like pairing
states. Apparently, the low temperature/energy behavior of
quasiparticle excitations is governed by the second band since the
first band is thermally activated. This would naturally explain why
the typical d-wave behaviors were observed in quite many experiments
\cite{sato01, armitage01, blumberg02, tsuei00b, chesca03}. However,
the presence of the first band will change the relative contribution
of the second band to the superfluid as well as other thermodynamic
functions. This will suppress, for example, the temperature
dependence of the normalized superfluid density and aggrandize the
experimental difficulty in identifying the expected power law
behavior for a d-wave superconductor.

The superfluid density is inversely proportional to the square of
the magnetic penetration depth, i.e., $\rho_s \propto \lambda^{-2}$.
Under the BCS mean-field decomposition, the superfluid density of
the system is simply a sum of the contribution from each band and
can be expressed as
\begin{equation}
\rho _{s}(T)=\rho _{s,1}(T)+\rho_{s,2}(T),
\end{equation}
where $\rho _{s,i}$ is the superfluid density of the {\it i}'th
band. It can be evaluated with the formula given in Ref.
\cite{xiang98}. In low temperatures, since there is a finite gap
in the quasiparticle excitations of the first band,
$\rho_{s,1}(T)$ is expected to be given by
\begin{equation} \rho_{s,1}(T)
\sim  \rho_{s,1}(0)\left( 1 - a e^{-\Delta^\prime_{1} /k_B
T}\right) ,
\end{equation}
where $\Delta_{1}^\prime$ is the minimum value of $\Delta_1
\gamma_k$ on the FS of the first band and $a$ is a constant. There
are gap nodes in the second band, therefore $\rho_{s,2}$ should
behave similarly as in a pure d-wave superconductor and show a
linear $T$ dependence in low temperatures due to the low energy
linear density of states:
\begin{equation}
\rho_{s,2}(T) \sim \rho_{s,2}(0)\left( 1 - \frac{T}{T_c}\right) .
\label{dwave}
\end{equation}
Thus, in the limit $T\ll T_c$, the normalized total superfluid
density is approximately given by
\begin{equation}
\frac{\rho_s (T)}{\rho_s(0)} \approx 1 -
\frac{\rho_{s,2}(0)}{\rho_{s}(0)}\frac{T}{T_c} -
\frac{\rho_{s,1}(0)}{\rho_{s}(0)}a e^{-\Delta^\prime_{1} /k_B
T}\label{mix}
\end{equation}
where $\rho_s (0) = \rho_{s,1}(0) + \rho_{s,2}(0)$.

For a pure d-wave superconductor, as shown by Eq. (\ref{dwave}),
the slope of the linear $T$ term in the normalized superfluid
density is proportional to $1/T_c$. However, for the coupled
two-band system considered here, this linear slope is normalized
by a factor $\rho_{s,2}(0) / \rho_{s}(0)$. The zero temperature
superfluid density $\rho_{s,i}(0)$ is a measure of the diamagnetic
response in the {\it i}'th band. It is approximately proportional
to the ratio between the charge carrier concentration and the
effective mass in that band, {\it i.e.}, $\rho_{s,i}(0)\propto n_i
/ m_i^*$. It is difficult to estimate this ratio for each
individual band. However, as the FS pocket of the first band
appears immediately after doping and that of the second band
appears only after the long range antiferromagnetic order is
completely suppressed, one would expect $\rho_{s,2}(0)$ to be much
smaller than $\rho_{s,1}(0)$. This means that $\rho_{s,2}(0) /
\rho_{s}(0) \ll 1 $ and the linear $T$ term in $\rho_s(T)$ is
greatly suppressed. Thus the low temperature curve of the
normalized superfluid density looks much flatter than in a pure
d-wave system, although $\rho_s(T)$ is still governed by a power
law $T$ dependence at sufficiently low temperatures.

\begin{figure}[h]
\includegraphics[width = 8cm]{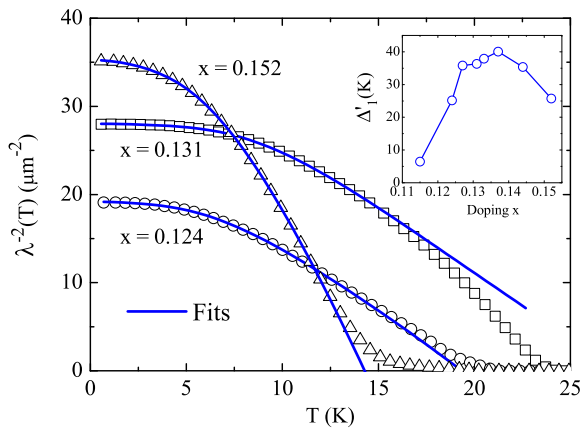}
\caption{Fitting curves of Eq. (\ref{sup1}) to the low temperature
superfluid density data published in Ref. [\onlinecite{kim03}] for x
= 0.124, 0.131 and 0.152. The inset shows the doping dependence of
the fitting parameter $\Delta_1^\prime$.} \label{fig1}
\end{figure}

In real materials, the low temperature dependence of ${\rho_s (T)} /
{\rho_s(0)}$ will be further suppressed by impurity scattering and
the linear term will be replaced by a $T^2$ term in the limit $T\ll
\Gamma_0$ \cite{Hirschfeld}
\begin{equation}
\rho_{s,2}(T) \sim \rho_{s,2}(0) \left(1 - \frac{k_B^2 T^2}{6\pi
\Gamma_0 \Delta_2} \right),
\end{equation}
where $\Gamma_0$ is the scattering rate. In this case, ${\rho_s (T)}
/ {\rho_s(0)}$ becomes
\begin{equation}
\frac{\rho_s (T)}{\rho_s(0)} \approx 1 -
\frac{\rho_{s,2}(0)}{\rho_{s}(0)}\frac{k_B^2 T^2}{6\pi \Gamma_0
\Delta_2}  - \frac{\rho_{s,1}(0)}{\rho_{s}(0)}a
e^{-\Delta^\prime_{1} /k_B T}. \label{sup1}
\end{equation}
We believe this formula captures the main feature of low temperature
superfluid density. Indeed, by fitting the experimental data with
the above equation, we find that it does give a good account for the
low temperature superfluid in the whole doping range. This can be
seen from Fig. \ref{fig1} where the fitting curves of Eq.
(\ref{sup1}) to the measurement data published in Ref. \cite{kim03}
are shown for three representative doping cases in the under-,
optimal and over-doping regimes, respectively.

In electron-doped materials, doping will reduce the distance
between the FS of the first band and the nodal lines of
$\gamma_k$. At low doping, the contribution from the exponential
term is small and the $T^2$ term is dominant. By further doping,
$\Delta_1^\prime$ begins to drop (the inset of Fig. \ref{fig1}),
the contribution from the exponential term becomes comparable with
the $T^2$ term in certain low temperature regime. In this case,
the $T^2$ dependence of ${\rho_s (T)}$ would become difficult to
be identified if the exponential term is not clearly separated. In
heavily overdoped regime, the FS of the first band will stride
over the nodal lines of $\gamma_k$. In this case, $\Delta_1^\prime
=0$ and $\rho_s(T) / \rho_s(0)$ should behave similarly as in a
conventional d-wave superconductor. This picture for the doping
dependence of low temperature $\rho_s(T) / \rho_s(0)$ agrees
qualitatively with all experimental observations.

\begin{figure}[ht]
\includegraphics[width = 7.5cm]{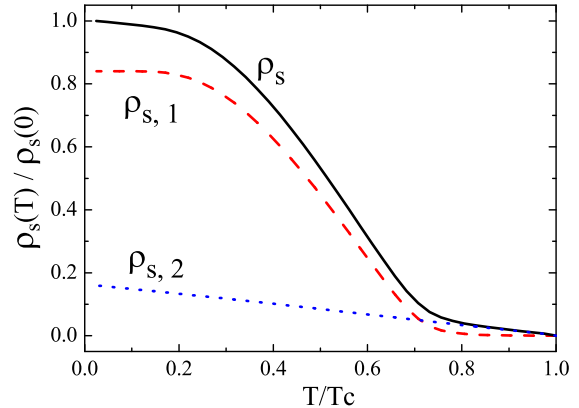}
\caption{Illustration of the contributions from the two bands to the
superfluid density in electron-doped cuprates. } \label{fig2}
\end{figure}

Close to $T_c$, a positive curvature will appear in $\rho_s(T)$.
This is a simple but universal property of a weakly coupled two-band
system\cite{xiang96}. To understand this, let us first consider the
case $g_3=0$. In this case, the two bands are decoupled and will
become superconducting independently. Let us denote their transition
temperatures by $T_{c1}^0$ and $T_{c2}^0$ and assume $T_{c1}^0 <
T_{c2}^0$. For finite but small $g_3$, the superconducting
transition will occur at a critical temperature close to $T_{c2}^0$,
i.e. $T_c \sim T_{c2}^0$ (Fig. \ref{fig2}). Just below $T_c$,
$\rho_s$ is mainly contributed from the second band. However, when
$T$ drops below $T_{c1}^0$, the intrinsic superconducting
correlation of the first band will appear in addition to the induced
one, and the contribution to $\rho_s(T)$ from this band will rise
rapidly with decreasing temperature. Consequently, a clear upturn
will show up in $\rho_s(T)$ around $T_{c1}^0$. The appearance of a
positive curvature in the experimental data of $\rho_s(T)$, as
already mentioned, is a strong support to the two-band picture.

To calculate explicitly the temperature dependence of $\rho_s$ in
the whole temperature range, one needs to know the band dispersion
$\xi_{ik}$. For this purpose, we adopt the expressions first
proposed by Kusko {\it et al.} \cite{kusko02} $ \xi _{ik} = $ $\pm
\left(\varepsilon _{i,k}+\varepsilon _{i,k+Q}\pm \sqrt{ (\varepsilon
_{i,k}-\varepsilon _{i,k+Q})^{2}+4\delta ^{2}}\right)/2-\mu _{i}$
where $\pm $ corresponds to the first/second band, $ \varepsilon
_{ik}=-2t_{i}(\cos k_{x}+\cos k_{y})-4t_{i}^{\prime }\cos k_{x}\cos
k_{y}-4t_{i}^{\prime \prime }(\cos ^{2}k_{x}+\cos ^{2}k_{y}-1)$,
$t_{i}^{\prime }=-0.25t_{i}$ and $t_{i}^{\prime \prime }=0.2t_{i}$.
$Q=(\pi ,\pi )$ is the antiferromagnetic wave vector and here
$\delta $ is taken as a constant. $\mu _{i}$ is the chemical
potential determined by the occupation number for each band. It was
shown that the FS contours determined from this formula agree
qualitatively with the ARPES data\cite{kusko02,yuan04}. Following
the suggestion of Ref. [\onlinecite{wang91, jiang94}], we assume
that the second band is hole-like. The doping concentration is
therefore given by the difference $x = n_e - n_h$, where $n_e$ and
$n_h$ are the carrier concentrations of the first and the second
bands, respectively. However, it should be emphasized that similar
results can also be obtained if both bands are electron-like.

\begin{figure}[tbp]
\includegraphics[width = 8cm]{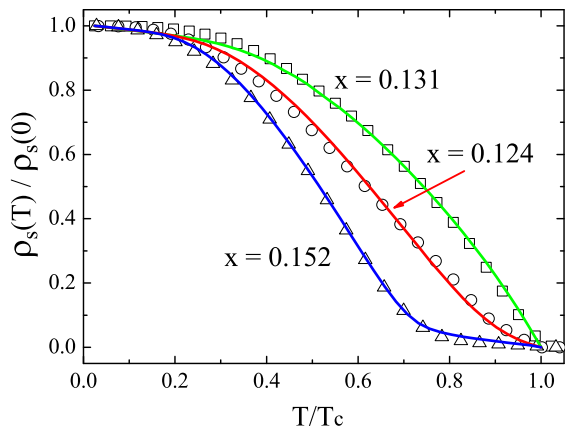}
\caption{Comparison between theoretical calculations (lines) and
experimental data (symbols) \protect\cite{kim03} for the temperature
dependence of the normalized superfluid density
$\rho_s(T)/\rho_s(0)$ of PCCO at three different doping levels.}
\label{fig3}
\end{figure}

Fig. \ref{fig3} compares the theoretical results of superfluid
density for a pure system with the corresponding experimental data
(symbols) \cite{kim03} for $x = 0.124$, 0.131, and 0.152. The
parameters used are $t_1 = 5$, $t_2 = 1$, $( g_{1},g_{2},
g_{3},n_{e}, n_{h} ) = ( 1.3,1.082, 0.005,0.214,0.09) $ for
$x=0.124$, $( 1.3,1,0.01, 0.231,0.1)$ for $x=0.131$, and $(
1.3,0.984,0.001, 0.261,0.11)$ for $x=0.152$. As can be seen, the
overall agreement between theoretical calculations and
experimental data is fairly good. It gives a strong support to our
picture. In low temperatures, the theoretical curves exhibit
stronger temperature dependence than the experimental ones,
especially for the case $x=0.131$. This is because the impurity
scattering was not included in the theoretical calculations. By
including the impurity scattering, the linear temperature behavior
$\rho_s$ will be replaced by a quadratic form. This, as
demonstrated in Fig. \ref{fig1}, will reduce the difference
between theoretical calculations and experimental data in low
temperatures.

Besides the superfluid density discussed above, our model is also
consistent with the phase-sensitive, tunneling spectroscopy, and
other experiments that support d-wave pairing symmetry in
electron-doped cuprates. Recently, the ARPES \cite{Matsui04} as
well as the Raman spectroscopy showed that the energy gap is
highly anisotropic and shows a maximum between the nodal and
anti-nodal regions. This non-monotonic variation of the energy gap
from the zone diagonal to the zone axis is not what one may expect
for a single-band d-wave superconductor, but is compatible with
our two-band picture.

More experimental measurements should be done to further detect the
gap structure in electron-doped materials. The scanning tunneling
measurement that was used for testing the two-band nature of MgB$_2$
from the vortex core state along the c-axis\cite{eskildsen02}, for
example, can be used to examine the two-gap picture here. Since the
interlayer hopping is highly anisotropic \cite{xiang98} and the
c-axis tunneling current is contributed mainly from the first band,
this measurement would allow us to determine the coherence length of
the first band from the spatial extension of the vortex core.
Comparing it with the coherence length of the second band which can
be determined from the measurement of $H_{c2}$, this will provide a
direct test for our two-band theory.

In conclusion, we showed that the temperature dependence of
$\rho_s$ in electron-doped cuprate superconductors can be well
explained by a weakly coupled two-band model. Our work resolves
the discrepancy in the interpretation of different measurement
results. It suggests that the pairing potential in electron-doped
cuprates has $d_{x^2-y^2}$ symmetry in the whole doping range,
same as in hole-doped materials.

We are grateful to T. R. Lemberger, D. H. Lee and L. Yu for helpful
discussions, and M. S. Kim and T. R. Lemberger for kindly providing
us the experimental data. This work was supported by the National
Natural Science Foundation of China.

\end{document}